\documentclass{pasj00}
\draft
\Received{Mar. 31, 2007}
\Accepted{Oct. 10, 2007, PASJ}
\begin{document}

\title{The 2006 Radio Outburst of a Microquasar Cyg X-3: Observation and Data}
\SetRunningHead{Tsuboi et al.}{Radio Outburst of Cyg X-3}
\author{M. T\sc{suboi}, T. T\sc{osaki}, N. K\sc{uno},  K. N\sc{akanishi}, T. S\sc{awada}, T. U\sc{memoto}}
\affil{Nobeyama Radio Observatory\thanks{The Nobeyama Radio Observatory is a branch of the National Astronomical
Observatory, National Institutes of Natural Sciences, Japan.}, Minamimaki, Minamisaku, Nagano, 384-1305}
\author{S.~A. T\sc{rushkin}}
\affil{Special Astrophysical Observatory RAS, Nizhnij Arkhyz, Karachaevo-Cherkassia 369167, Russia }
\author{T. K\sc{otani}, N. K\sc{awai}}
\affil{Tokyo Tech, 2-12-1 O-okayama, Meguro, Tokyo 152-8551}
\author{Y. K\sc{urono}, T. H\sc{anda}, K. K\sc{ohno}}
\affil{Institute of Astronomy, The University of Tokyo, Mitaka, Tokyo 181-0015}
\author{T. T\sc{sukagoshi}}
\affil{The Graduate University for Advanced Studies, 2-21-1 Osawa, Mitaka, Tokyo 181-0015}
\author{O. K\sc{ameya}, H. K\sc{obayashi}}
\affil{Mizusawa VERA Observatory, Mizusawa, Oshu, Iwate 023-0861}
\author{K. F\sc{ujisawa}, A. D\sc{oi}}
\affil{Faculty of Science, Yamaguchi University, Yamaguchi, Yamaguchi 753-8512}
\author{T. O\sc{modaka}}
\affil{Faculty of Science, Kagoshima University, Kagoshima, Kagoshima 890-0065}
\author{H. T\sc{akaba}, H. S\sc{udou}, K. W\sc{akamatsu}}
\affil{Faculty of Engineering, Gifu University, Gifu 501-1193}
\author{Y. K\sc{oyama}, E. K\sc{awai}}
\affil{ National Institute of Information and Communications Technology, Kashima, Ibaraki 314-8501}
\and
\author{N. M\sc{ochizuki}, Y. M\sc{urata}}
\affil{Institute of Space and Astronautical Science, Sagamihara, Kanagawa 229-8510}

\KeyWords{black hole physics --- stars: variables: other --- radio continuum: stars
}
\maketitle

\begin{abstract}
We present the results of the multi-frequency observations of radio outburst of the
microquasar Cyg X-3 in February and March 2006 with the Nobeyama 45-m
telescope, the Nobeyama Millimeter Array, and the Yamaguchi 32-m
telescope.  Since the prediction of a flare by RATAN-600,
the source has been monitored from Jan 27 (UT) with these radio
telescopes.   At the eighteenth day after the quench of the activity, successive flares
exceeding 1 Jy were observed successfully.
The time scale of the variability in the active phase is presumably
shorter in higher frequency bands.

We also present the result of a follow-up VLBI observation at 8.4 GHz with the Japanese VLBI Network (JVN) 2.6 days after the first rise. The VLBI
image exhibits a single core with a size of $<8$ mas (80 AU). The
observed image was almost stable, although the core showed rapid
variation in flux density.  No jet structure was seen at a sensitivity of
$T_b = 7.5\times 10^5$ K\@.
\end{abstract}

\section{Introduction}
Cyg X-3 is a famous X-ray binary including a black hole candidate (e.g., \cite{schalinski1998}). This object is classified as a microquasar due to its bipolar relativistic jet accompanied by radio flares.  Because it is located on the Galactic plane at a distance of about 10 kpc (e.g.,
\cite{predehl2000}) and obscured by intervening interstellar matter, it has been observed mainly in radio and X-ray.  Its giant radio flares have been observed once every several
years since its initial discovery \citep{gregory1972,braes1972}.  The peak flux densities in the radio flares have often increased up to levels of 10 Jy or more at centimeter wave (e.g.,
\cite{waltman1994}). The radio emission seems to be correlated with hard X-ray emission, and not with soft X-ray emission \citep{mccollough1999}. Although the radio emission arises through synchrotron process of relativistic electrons in the jet \citep{hjellming1988}, the millimeter behavior during the flares is not yet established.  An observation at a shorter wavelength and with a higher time resolution is desirable to understand the mechanism of the flares.

The quenched state of Cyg X-3, in which the radio emission is suppressed
below 1 mJy, is a possible precursor of flares (e.g.,
\cite{waltman1994}).  In January 2006, this quenched state was
detected in monitoring observations with the RATAN-600 radio telescope
\citep{trushkin2006}.  The source has been monitored from MJD$=53762$
(Jan 27 2006 in UT) with the Nobeyama 45-m radio telescope (NRO45), the
Nobeyama Millimeter Array (NMA), and the Yamaguchi 32-m radio telescope
(YT32).  We detected the initial state, or  rising phase, of the
radio flare of Cyg X-3 at MJD$= 53768$ (February 2 2006) and observed
successive flares exceeding 1 Jy \citep{tsuboi2006}, which turned out
to be the beginning of an active phase lasting more than 40 days .

In this paper, radio observations with NRO45, NMA, YT32, and the
Japanese VLBI Network (JVN) are reported.  The observation procedures
are summarized in section 2. The light curves and the spectral
evolution observed with NRO45, NMA and YT32 are shown in section 3,
together with the result of JVN\@.  Detailed discussion based on
these observations will be published as separate papers.

\section{Observations and data reductions}
\subsection{Radio photometric observations}
The first observation period was from MJD$= 53763.13$ (January 28 2006) to
53779.94 (February 13 2006).  Observations with NRO45 of Cyg X-3 were
 performed alternately at 23 GHz and at both 43 and 86 GHz,
simultaneously.  The period corresponded to the initial phase of the
radio flaring state in February-March 2006.  The second period was from
MJD$=53805.12$ (2006 March 11) to $53808.77$ (2006 March 14).  A cooled HEMT
receiver with dual circular polarization feed was used at 23 GHz.
SIS receivers with orthogonal linear polarization feeds were used
at 43 and 86 GHz.  The system noise temperatures during the
observations, including atmospheric effects and antenna ohmic loss, were
80--120 K at 23 GHz, 120--200 K at 43 GHz, and 250--350 K at 86 GHz.  The full width at half maximums (FWHM) of the telescope
beams are $77''$ at 23 GHz, $39''$ at 43 GHz, and $19''$ at 86 GHz.  The telescope beam was alternated between the positions of
the source and sky at 15 Hz by the beam-switch in order to subtract
atmospheric effect.  Antenna temperatures were calibrated by the
chopper wheel method.  The primary flux calibrator for conversion from
antenna temperature to flux density was a proto-planetary nebula,
NGC 7027, whose flux density values are given as 5.5 Jy at 23 GHz, 5.0
Jy at 43 GHz, and 4.6 Jy at 86 GHz \citep{ott1994}.  Telescope pointing
was checked and corrected in every observation procedure by
observing NGC 7027 in cross-scan mode.  The pointing accuracy was
better than $3''$ r.m.s. during these observations.  The source was
observed using ON-OFF observations of durations of 5--10 minutes,
sufficient to detect and perform photometry on Cyg X-3 and the
calibrator.

Interferometric observations were performed with the NMA from MJD$=
53762.18$ (January 27 2006) to $53776.15$ (February 10 2006) at both 98 and
110 GHz simultaneously.  The NMA consists of six 10m antennas equipped
with cooled DSB SIS receivers with a single linear polarization
feed. The Ultra-Wide-Band Correlator with a 1GHz bandwidth was
employed for the backend \citep{okumura2000}. The quasar 2017+370 was
used as a phase and amplitude reference calibrator and Uranus and
Neptune were used as primary flux-scale calibrators. The system noise
temperatures during the observations, including atmospheric effects and
antenna ohmic loss, were 80--120 K at 98 GHz and 120--200 K at 110 GHz. The uv-data were calibrated with the UVPROC-II
software package developed at NRO \citep{tsutsumi1997}, and then imaged
with natural UV weighting, and CLEANed with the NRAO AIPS package.

Centimeter-wave observations of Cyg X-3 were also performed at 8.4
GHz for longer duration with YT32.  The observation period was from
MJD$=53768.29$ (February 2 2006) to $53813.08$ (March 19 2006). A cooled
HEMT receiver was used at 8.4 GHz. The system noise temperature of YT32
during the observations, including atmospheric effects and antenna ohmic
loss, was 45 K at 8.4 GHz. The primary flux calibrator for YT32 was an H
{\sc ii} region, DR21 with a flux density of 20 Jy at 8.4 GHz. Flux
measurement was carried out with ON-OFF switching method with
an overlaying small-angle offset for both azimuth and elevation directions.
 Additional observations using YT32 were also  performed in
May 2006. Data at MJD$=53872$ was obtained with the Mizusawa VERA
Observatory 10-m radio telescope.

The uncertainty in flux density of Cyg X-3 depends on weather
conditions. However, sensitivity of telescopes is not the principal factor of the uncertainty.  Because the primary flux-scale calibrator for NRO45, NGC 7027, is close to Cyg X-3 in the celestial sphere, the difference of atmospheric attenuation between these sources has no significant effect on the data.  The typical systematic uncertainty is $\sim10$\% for NRO45. The typical systematic uncertainty of NMA is $\sim15$\% because the primary flux-scale calibrators are not near to Cyg X-3 and phase noise caused by atmospheric fluctuations. Although flux loss due to pointing errors is corrected by  pointing offset data in the data reduction process, the uncertainty  of YT32 is  as much as $\sim20$\%.  However, the relative uncertainties of  flux density in a day should be much better than these values.

\subsection{JVN observation}
A follow-up VLBI observation was carried out with the Japanese VLBI Network
(JVN; \cite{fujisawa2007,doi2006a,doi2006b}). The duration of the
observation was from MJD$=53770.8$ to MJD$=53771.4$, i.e., starting 2.6 days
after the first rise of the flare.  The telescopes participating in this
observation are four 20-m telescopes of the VLBI Exploration of Radio
Astrometry project (VERA; \cite{kobayashi2003}), Usuda 64-m (U64),
Kashima 34-m (K34), YT32, and Gifu 11-m (G11).  Right-circular
polarization was received at 8400-8416 MHz (IF1) and 8432--8448 MHz
(IF2) with a total bandwidth of 32 MHz.  The VSOP/K4-terminal system was
used as a digital back-end; digitized data in 2-bit quantization are
recorded onto magnetic tapes at a data rate of 128 Mbps.  Two sources
(2000+472, 3C454.3) other than Cyg X-3 were observed for gain and
bandpass calibration, respectively.  The data were correlated with the
VSOP-FX correlator at NAOJ (\cite{shibata1998}), and fringe were detected
at all baselines except for baselines including Ishigaki, Ogasawara, and
G11 telescopes.

The data was reduced in the standard manner with the Astronomical
Image Processing System (AIPS; \cite{greisen2003}) developed at the US
National Radio Astronomy Observatory.  An amplitude-scaling factor was
determined from monitored system noise temperatures and antenna
efficiencies of U64, YT32, and K34 telescopes.  Furthermore, we 
calibrated antenna gain variations using the data of 2000+472, which is
a point source in the JVN baselines and scanned every 30--60 minutes.
Such a calibration method provides an absolute flux scale with an
accuracy of $\sim10$\% and a relative time variation of antenna gain
of $\sim3$\% accuracy.

\section{Results}
\subsection{Light curves}
We present the light curves of Cyg X-3 at 6 frequency bands of 8.4, 23,
43, 86, 98, and 110 GHz obtained with NRO45, NMA, and YT32 in Fig. 1.
Table 1 summarizes results of observations.  The first rise of the flare
was detected with NMA at MJD$= 53768.13$ or 3 am on February 2 2006 (UT)
at 98 and 110 GHz. This is about 18 days after it entered the quenched
state observed with RATAN-600.  The first rise was also observed at
lower frequencies within 1 day.   After that, we observed several peaks
exceeding 1 Jy.  Although there is a long intermission of observation in
our campaign, the duration of the active phase of Cyg X-3 is at least
over 40 days \citep{tosaki2006}.  It can be confirmed that Cyg X-3 was
still active in May from the flux densities at 8.4, 98, and 110 GHz in
2006 May in Fig. 1 and Table 1.

\begin{figure}
\centering
\includegraphics[height=15.5cm]{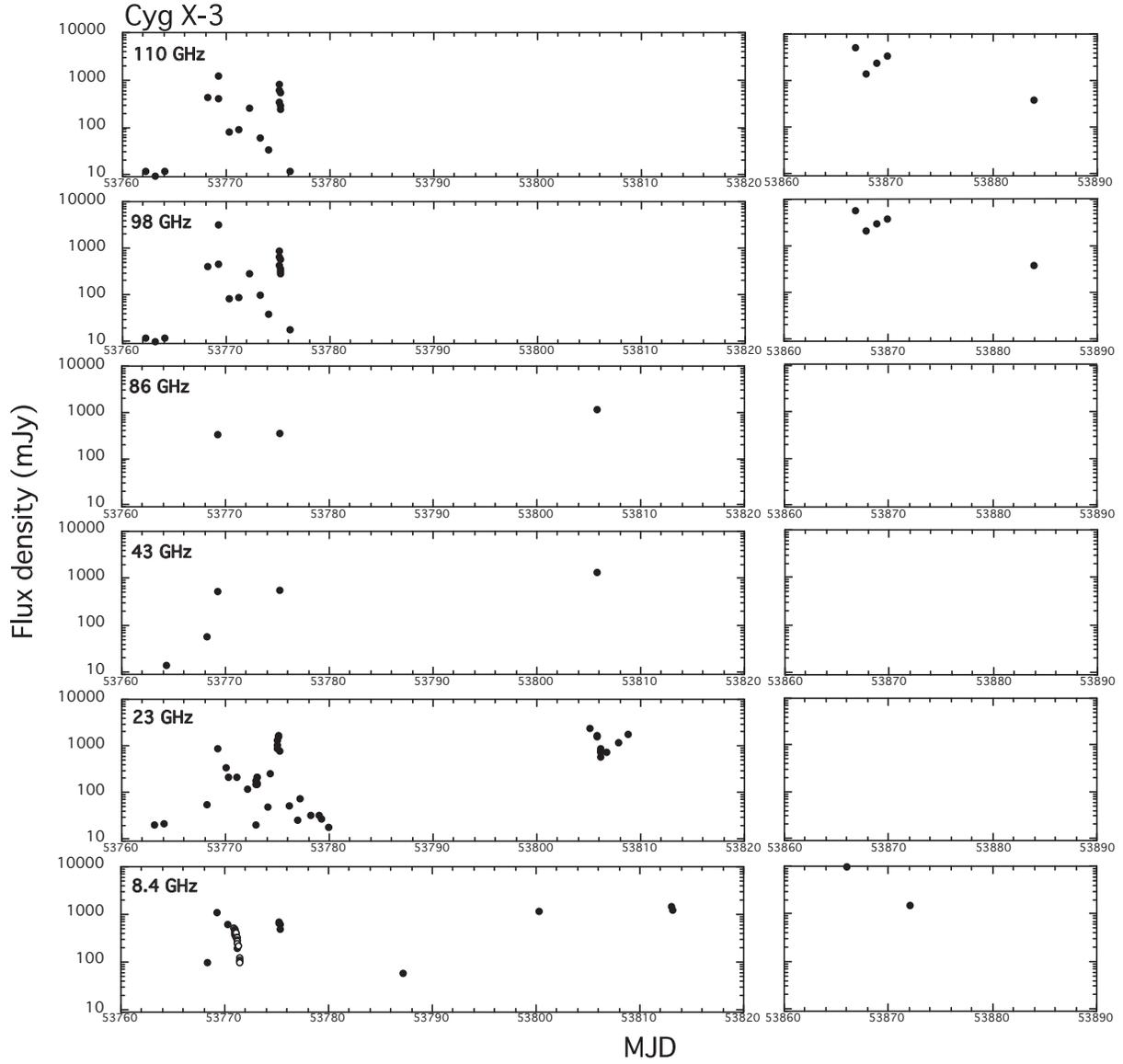}
\label{fig.1a}

\caption{Radio light curves of February to May 2006 of Cygnus X-3 at,
 from bottom to top, 8.4, 23, 43, 86, 98, and 110 GHz. The 8.4 GHz data 
were obtained with YT32.  In addition, open circles at 8.4 GHz show high-density sampling flux 
densities observed by the Japanese VLBI Network (JVN). The 23, 43 and 
86 GHz data were obtained with NRO45.  The 98 and 110 GHz data were 
obtained with NMA. The first rise was detected at MJD$= 53768.13$
  at 98 and 110 GHz. Within one day after the first rise, the flux density at 23 GHz also
increased. The 8.4, 98, and 110 GHz data show Cyg X-3 was in an active phase in May 2006. }
\end{figure}

Figure 2a shows the enlarged light curves of the initial phase of the
first flare.  Before the flare, the flux density of Cyg X-3 was
inhibited up to a few 10 mJy at 23 to 110 GHz (see Figure 1).  At
MJD$= 53768.13$, the first rise was found by NMA at 98 and 110 GHz,
of which the flux densities became 40 times or more compared with the value of the previous day. The
flux densities at 23 and 43 GHz with NRO45 were also 3-4 times brighter
than the previous values. The peak flux densities at MJD$= 53769.17$
exceed 3 Jy at 98 GHz and 1 Jy at 110 GHz.  They were
violently variable and decreased to 1 Jy or less within one hour.
Assuming an exponential decay, the e-folding decay times would be 0.03
days both at 98 GHz and 110 GHz.  On the other hand, the e-folding decay
time at 8.4 GHz are $t= 0.36 \pm 0.02$ day during MJD$=53770$ to $53771$,
which will be mentioned for details.  For several major flares of Cyg
X-3 observed previously, the e-folding decay time of the flux density was
reported to be in the range of 0.15 to 2.75 days \citep{hjellming1974}.  The observed
decay times are much shorter than these previous values.

\begin{figure}
\centering
\includegraphics[width=15.5cm,clip]{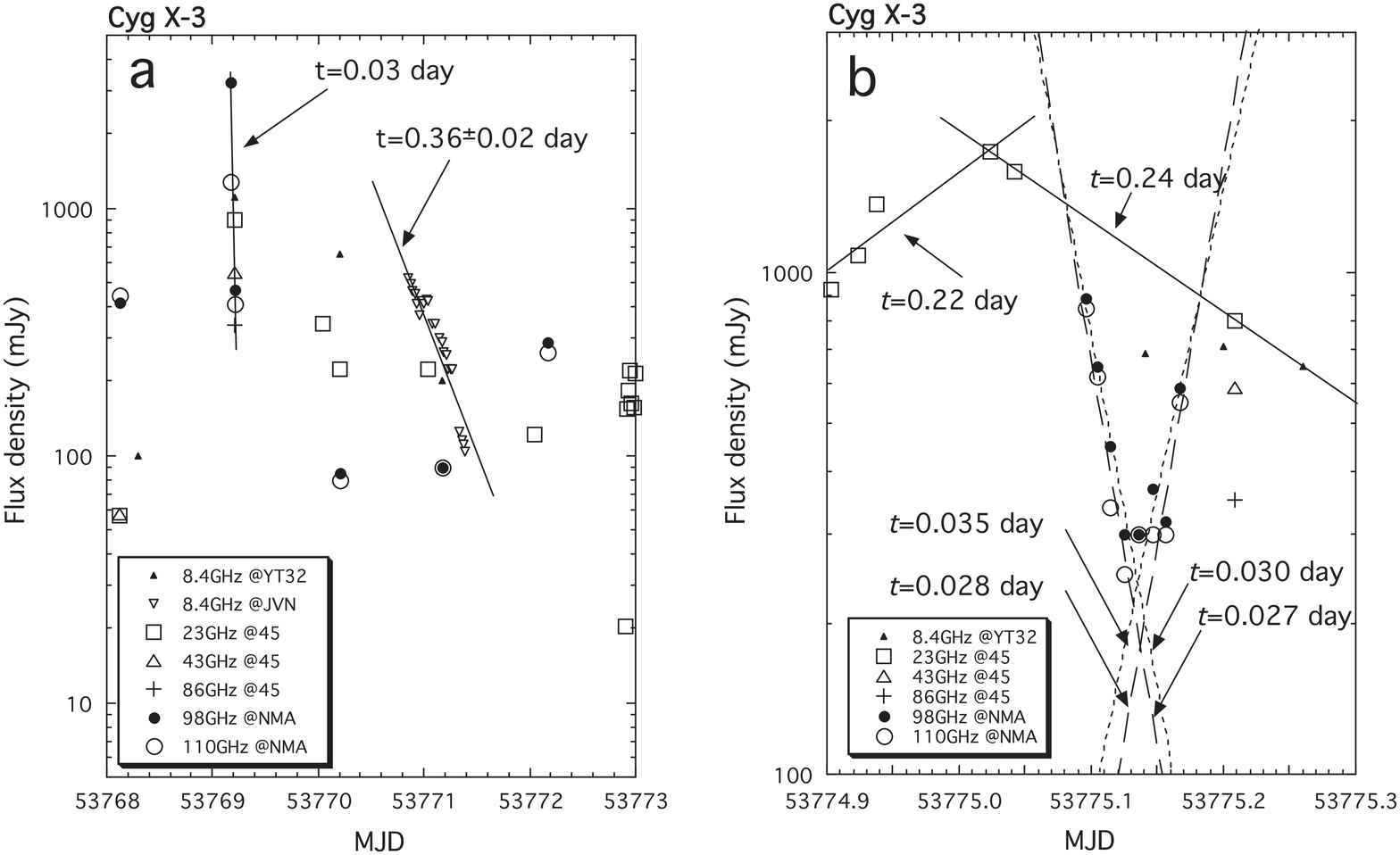}
\label{fig.2}

\caption{Enlarged radio light curves of Cyg X-3 at the first and second
flares.

 {\bf (a)}~Observation epoch is from MJD = 53768 to 53773. 
At MJD$= 53768.13$, the first rise was found by NMA at 98 and 110 GHz, and
the flux densities increased up to 0.4 Jy. This is 40 times or more compared
 with the value before the first rise. The flux densities at MJD$= 53769.17$
exceed 3 Jy at 98 GHz and 1 Jy at 110 GHz.  They 
decreased to 1 Jy or less within one hour.   The best-fitting curves of
  exponential decay model are plotted in lines. Open upside-down triangles
   show flux densities at 8.4 GHz observed by the
Japanese VLBI Network (JVN). The flux densities at lower 
 frequencies changed more gradually.

{\bf(b)}~Observation epoch is from MJD $= 53774.9$ to $53775.3$.  Data with NMA at 
98 and 110 GHz show clearly the quench and the successive rapid rise 
of the millimeter flux. The best-fit curves of exponential rise and decay models 
are plotted in lines. The e-folding rise and decay times are $t\simeq 0.03$ day both at 98 and 110 GH. }
\end{figure}

Figure 2b shows the enlarged light curves of the second flare.  We
obtained a higher sampling rate light curve at 23 GHz in the rising phase
of the second flare. At MJD$=53774.9$, the rise of the second flare was
detected at 23 GHz with the NRO45. The flux density at 23 GHz
increased rapidly from 0.9 to 1.7 Jy within 3 hours. That corresponds to
an e-folding rise time of $t= 0.22 \pm 0.05$ day. The flux density
decreased rapidly from the peak to 0.8 Jy by the next observation 4.4
hours after. If these flux densities are involved in the same flare, the
e-folding decay time is $t= 0.24 \pm 0.01$ day. The decay phase of the
flare was also observed at other frequencies. The flux densities
observed at 98 and 110 GHz decreased rapidly within the observation
interval as at 23 GHz. A power law describes the decay behavior. The
e-folding decay times at 98 and 110 GHz are $t= 0.030 \pm 0.003$ day and
$t= 0.027 \pm 0.005$ day, respectively.  These are of the same order of
the first flare at 98 and 110 GHz. In addition to the peak and following
decay at 23 GHz, another rising is recognized at 98 and 110 GHz in the
observation break at 23 GHz. The e-folding rise times at 98 and 110 GHz
are $t= 0.035 \pm 0.025$ day and $t= 0.028 \pm 0.013$ day, respectively. The
e-folding rise times are also at the same level as the decay times.

Figure 3 shows the relation between the time scale of the flux variability
and frequency in the 2006 flares of Cyg X-3.  The e-folding rise and
decay times are derived from neighboring observations within 8
hours. Our data with higher sampling rates than previous reports should
reveal new information about the synchrotron jets of Cyg X-3.  The
e-folding time is shorter at higher frequencies, and we fit the trend
with a power-law model, shown as a straight line in the figure.
Although the scattering of the data points is large, the e-folding time
is inversely proportional to the observed frequency, $T=3.8f^{-1.03}$.
That cannot be explained in terms of the synchrotron bubble model,
which predicts the same time scale for all the frequencies.  It is
suggested that the injection and loss of energy of relativistic
electrons in the jet plays an important role in the spectral evolution
of the flares of Cyg X-3.

\begin{figure}
\centering
\includegraphics[width=10cm,clip]{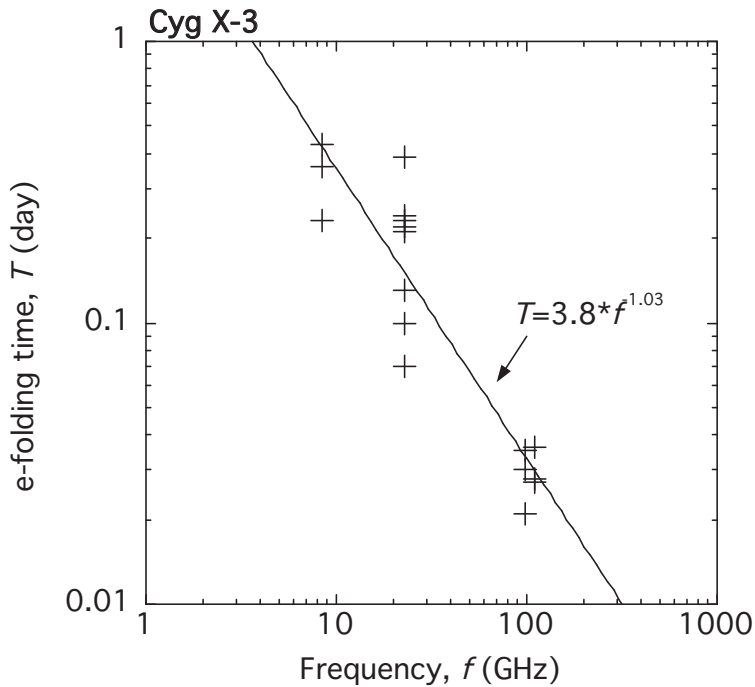}
\label{fig:3}
\caption{ The relation between the time scale of the flux variability and
frequency in the 2006 flares of Cyg X-3.  The solid line is the best-fitting
power-law model for these data. The e-folding time of the flux variability should  inversely relate frequency. }
\end{figure}

\begin{figure}
\centering
\includegraphics[width=15.5cm,clip]{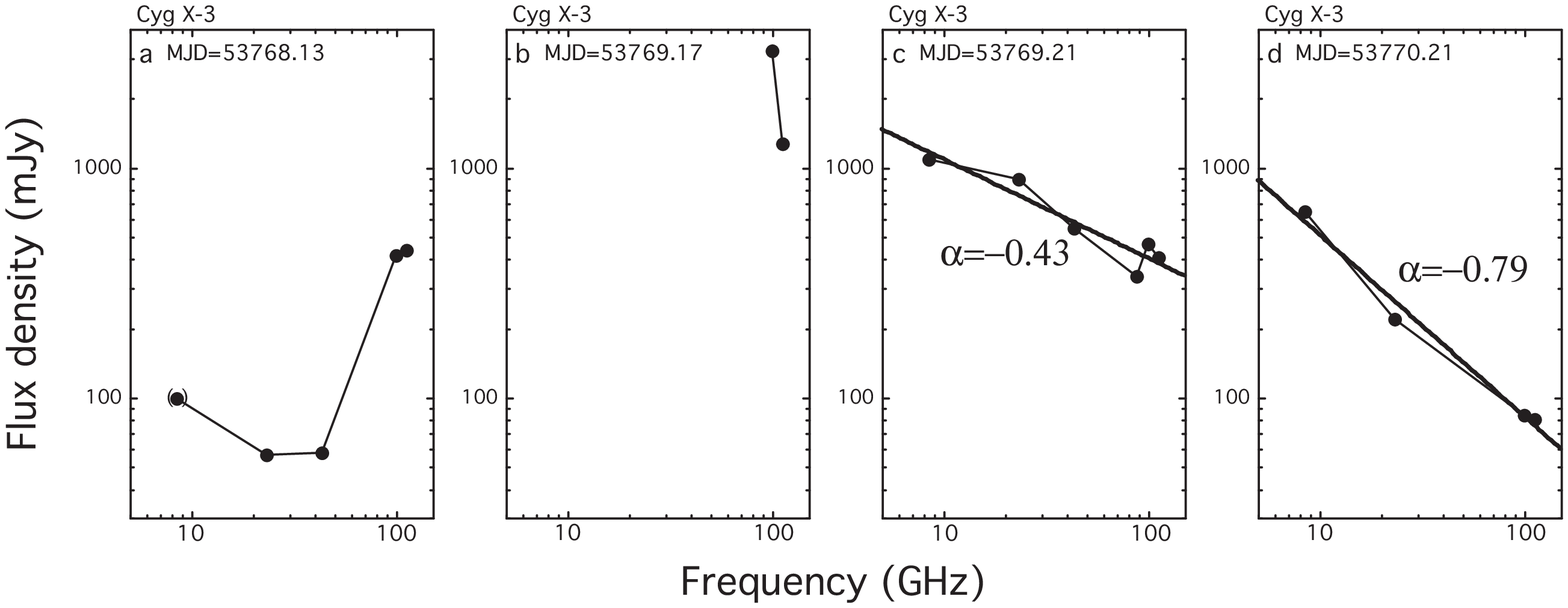}
\label{fig:4}
\caption{Spectral evolution during the first flare. Three days evolution around the first peak at MJD= 53768. 
  {\bf a}~On the first day at the first flare, MJD$= 53768.14$, the flux densities at 98 and 110 GHz increase rapidly, while the flux densities at lower frequencies were still low. 

{\bf b}~ 98 and 110 GHz data at MJD$=53769.17$.  The flux densities at 98
GHz and 110 GHz were 3.2 and 1.3 Jy, respectively.  These are the highest values in this flare. 
Unfortunately, it was not observed during this time at other frequencies. 

{\bf c} ~Flux densities after 1 hour of panel {\bf b}.  The flux densities at 8.4,22, and 43 GHz increased about 10 times over values in panel {\bf a}. Flux densities at 98 and 110 GHz decreased dramatically in 1 hour.  A power-law model, $S\propto f ^{\alpha}$, is applied to these (solid line).  The best fitting of the spectral index is $\alpha\simeq-0.4$.

{\bf d} ~Flux densities after 1 day of panel {\bf c}. Although the flux densities were described by a power law (solid line), the spectrum was fairly steepened. The best fitting of the spectral index is $\alpha\simeq-0.8$. }

\hspace{2cm}
\centering
\includegraphics[width=15.5cm,clip]{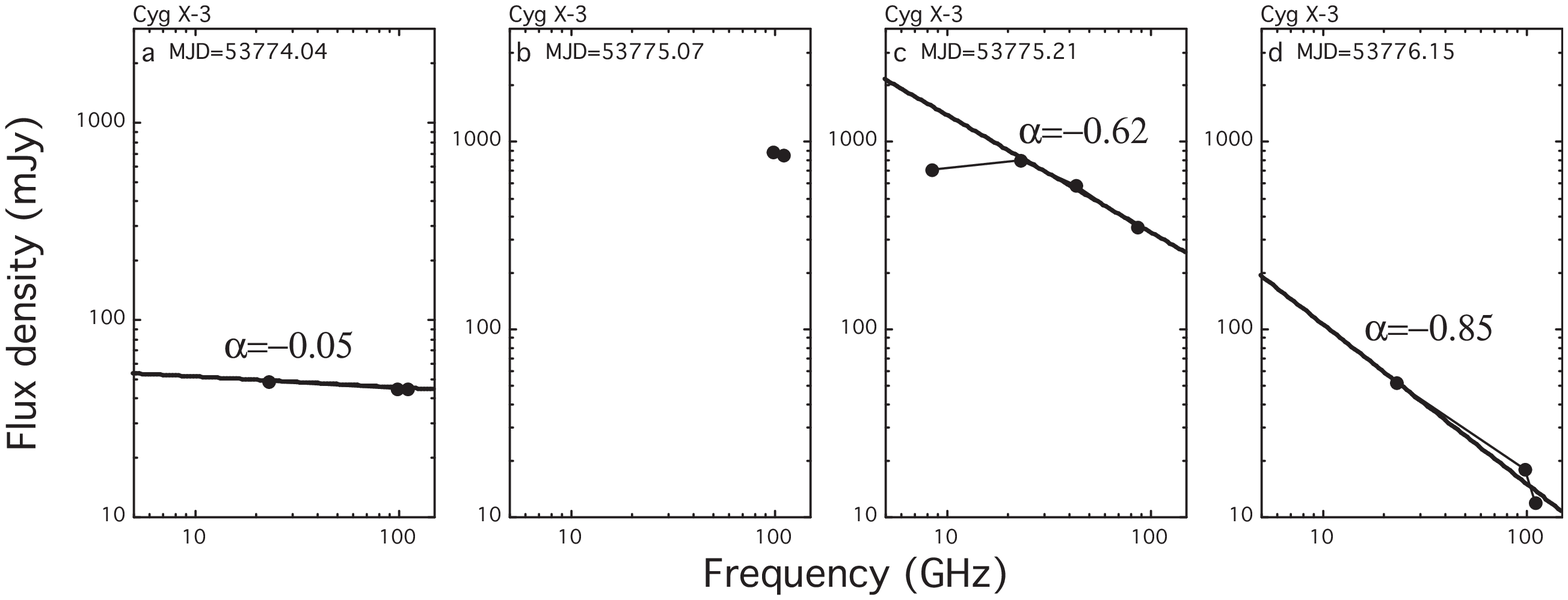}
\label{fig:5}
\caption{Three days spectral evolution during the second flare around MJD$= 53775$.  
{\bf a}~ Flux densities one day before the second flare,  or at MJD= 53774.04. The flux densities were described by a power law with $\alpha\simeq-0.1$ (solid line).

{\bf b}~Flux densities at MJD$= 53775.07$. These are the highest values at 98 and 110 GHz in this flare.  They would be near the peak of this flare because the highest value appeared at 22 GHz before 1 hour.  Unfortunately, it was not observed during this time at 98 and 110 GHz.

{\bf c} ~Flux densities after 3 hour of panel {\bf b}, Between these two panels, Cyg X-3 decreased and increased rapidly. (see also Fig 2b).  Flux densities except for at 8.4 GHz were described by a power law with $\alpha\simeq-0.6$ (solid line).

{\bf d} ~Flux densities after 1 day of panel {\bf c}.  The spectrum was steepened.  The best fitting of the spectral index is $\alpha\simeq-0.9$.}
\end{figure}

\subsection{Spectral Evolution of Flares}
Fig. 4 shows spectral evolution in the three days around the first
flare of Cyg X-3. The curve in panel {\bf a} indicates the spectrum at MJD$= 53768.13$, at the onset
of the first flare. The rapid increase at 98 and 110 GHz takes place
first as mentioned in the previous subsection, while the flux densities
at lower frequencies do not change significantly.  The spectrum shows a
complicated inverted feature from 43 to 98 GHz.  This suggests there is time lag of flare onset depending on frequency. 
The data in panel {\bf b}  indicate flux
densities at 98 and 110 GHz at MJD$=53769.17$.  The flux densities at 98
GHz and 110 GHz rise to 3.2 and 1.3 Jy, respectively.  These are the highest values in this flare. Unfortunately,  no simultaneous data at other frequencies are available.
Panel {\bf c} shows the  flux densities of Cyg X-3 at MJD$= 53769.21$, just after 1 hour of panel {\bf b}. The flux densities at lower frequencies increased about 10 times over values in panel {\bf a}, but flux densities at 98 and 110 GHz decreased dramatically in 1 hour.  Then the spectrum is described by a power law with an index
of $\alpha\simeq-0.4$.  The peak flux is followed by a rapid decay at high frequency.
Panel {\bf d} shows the spectrum at MJD$= 53770.21$, or 1 day after the
peak. The spectrum is also describable by a power law, but it was decayed to be as steep as $\alpha\simeq-0.8$. 
This is consistent with the decay time being shorter at higher frequency, as mentioned in the previous subsection. We assume that the spectral break seen at $\sim 100$ GHz at MJD$=53768.13$ moves in the spectrum down to below 8 GHz in one day. This evolution of the radio spectrum may be interpreted as the result of an adiabatic expansion of synchrotron emitting ejecta in Cyg X-3.

Figure 5 shows the spectral evolution in the three days around the
second flare. Panel {\bf a} in this figure indicates spectra on the day before the second peak or at MJD= 53774.04. This shows a flat spectrum with an index of $-0.1$, which suggests that the second flare has already started.  The spectrum at another flare in March was also flat over the observation band, suggesting an optically thick radio source.  A power law with an index of $-0.2$ can explain the spectrum. Panel {\bf b} shows flux densities at MJD$= 53775.07$.  These are the highest values at 98 and 110 GHz in this flare.  Following this, these decreased rapidly.  However, they would be near the peak of this flare, because the value at 22 GHz reached the maximum before 1 hour.  Unfortunately, it was not observed during this time at 98 and 110 GHz.
Panel {\bf c} shows flux densities after 3 hour of panel {\bf b}, Between these two panels, Cyg X-3 decreased and increased rapidly. (see also Fig 2b).  Flux densities except for at 8.4 GHz were described by a power law with $\alpha\simeq-0.6$ (solid line).  The spectral index is slightly steeper than that in the corresponding phase of the first flare. 
And panel {\bf d} shows flux densities after 1 day of panel {\bf c}.  The spectrum was steepened.  The best fitting of the spectral index is $\alpha\simeq-0.9$. The spectral index of $-0.9$ is similar to that one day after the first flare. The decay time of the second flare is also shorter at higher
frequencies, which cannot be explained by the sole synchrotron bubble model.  

\subsection{8.4 GHz VLBI results in a flare}
Our VLBI observation started from just 2~days after the first rise of
the flare, i.e., the first day after the intensity maximum of the first
flare.  Successive snapshots with a short integration time have
been made because rapid flux decrease has also been observed during the
VLBI observation.  Fig. 6 is an example of such a snapshot, of
which observation period is from MJD= 53770.979 to
53771.042.  Only a featureless (extended) structure is found in all snapshot
images.  Deconvolution with the structure model of an elliptical
Gaussian profile and self-calibration was done using DIFMAP software
\citep{shepherd1997}. The deconvolved source size was $\sim8$~mas.  This
is significantly broader than a synthesized beam size, 16.3~mas
$\times$ 4.3~mas at PA of $-2.9^\circ$.  However, no other component is
found beyond three times the r.m.s. of noise of the resultant image or
a brightness temperature of 7.5$\times$10$^{5}$~K\@.

\begin{figure}
\begin{center}
\includegraphics[width=8cm, clip]{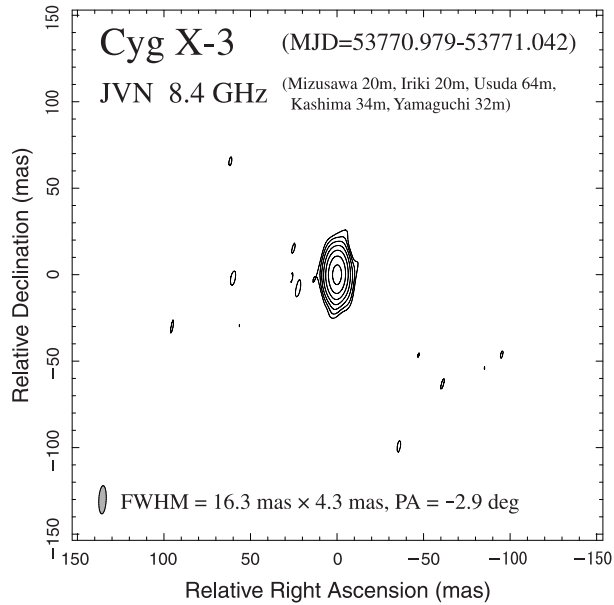}
\end{center}
\caption{Snapshot image of Cyg~X-3 from VLBI data in MJD $= 53770.979$ to
$53771.042$ (1.5~hour).  The beam size is 16.3~mas $\times$ 4.3~mas at PA
of $-2.9^\circ$, corresponding to 157~AU $\times$ 41~AU at the distance
to Cyg~X-3.  R.M.S.~of image noise, $\sigma$, is 0.70~mJy~beam$^{-1}$.
Contour levels are $3\sigma\times$(-1, 1, 2, 4, 8, 16, 32, 64).}

\label{fig:6}
\end{figure}

\begin{figure}
\centering
\includegraphics[width=8cm,clip]{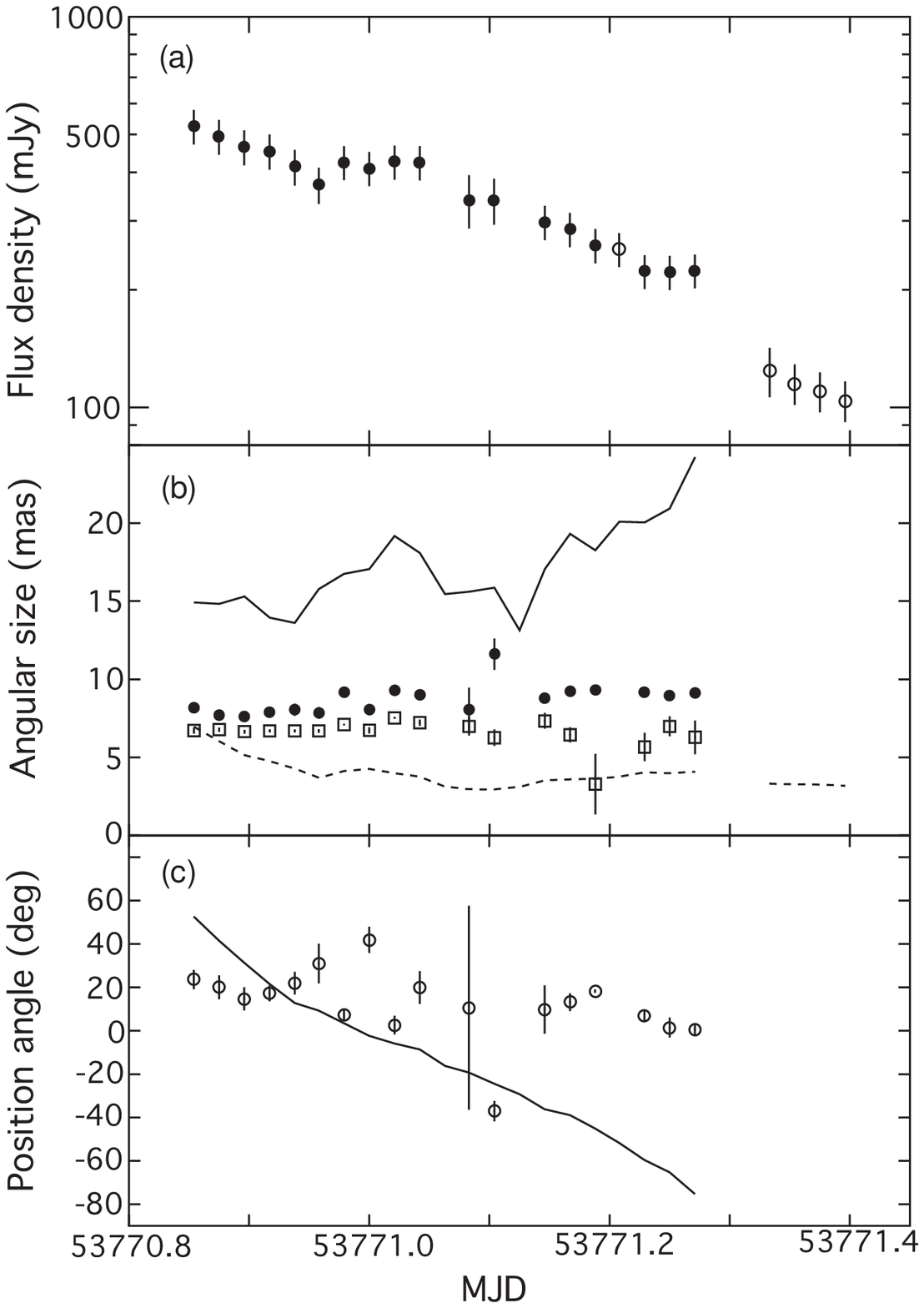}
\label{fig:7}

\caption{Evolution of source structure of Cyg~X-3 obtained with VLBI.
A series of measurements were performed by visibility-based model-fitting using 
an elliptical-Gaussian profile model with free parameters of flux density, 
major and minor axes of FWHM, and position angle of the major axis (see text in 
detail).  {\bf (a)}~Flux densities of VLBI component~(filled circle).  Open circles
represent measurements using a circular-Gaussian model, rather than
elliptical one, due to poor data quality.  {\bf (b)}~Angular sizes of
fitted source structure.  Filled circles and open squares represent the
FWHMs of major and minor axes, respectively.  Solid and dashed lines
represent the major and minor axes of HPBWs of synthesized beam in
uniform-weighting, for comparison.  {\bf (c)}~Position angles of major
axis of fitted source structure~(open circle).  Solid line represents
the position angle of the major axis of synthesized beam, for
comparison.}
\end{figure}

We also analyzed the time evolution of flux density and source structure during 
the JVN observation.  We have clipped out a series of data segments.  Each segment has a duration of 1~hour, and the interval of the starting time of the segments is set to be 0.5 hour, i.e., half of a duration overlapping the half of the next 
duration.  Using the task of UVFIT in AIPS, we measured a source size and a 
flux density for each clipped data by visibility-based model-fitting using an 
elliptical Gaussian profile model.   The flux density of the component  decreased from $\sim$500~mJy to $\sim$100~mJy during the
VLBI observation lasting $\sim$14~hours (Fig.2a and Fig 7a).  The source
structure has hardly changed, regardless of the variability
of the synthesized beam (Fig.7b and 7c).  The weighted averages of major
axis, minor axis, and position angles of the source structure are
8.9$\pm$0.1~mas, 7.1$\pm$0.1~mas, and 15.0 $\pm$2.3~deg, respectively,
i.e., we have resolved the source which is slightly but significantly
elongated in a north-south direction.  Our VLBI observation has revealed
that Cyg~X-3 has no significant structure change in spite of the rapid
flux variability at 8.4~GHz during the period.

Previous studies \citep{miller-jones2004} show that the two-sided jets
are ejected toward north and south after major radio flares. If the
ejection was concurrent with the first rise of the flux and the jet axis
is fairly close to perpendicular to the line of sight, the traveling
distance of the ejecta is expected to be 350~AU or 35~mas at 10~kpc.
However, we have not found any evidence of such a jet structure.  The
source size in our VLBI image is presumably affected by interstellar
scattering because it is consistent with the expected scattering size at
8.4~GHz \citep{schalinski1995}. Thus, there is no structural evolution
on a scale of larger than 5 mas in spite of rapid flux variability.
That may be a strong constraint on  jet evolution in a few days after
a radio flare.

\bigskip 
The authors would like to thank the members of NRO45 group and
NMA group of Nobeyama Radio Observatory for support in the
observations. This work is partially supported by the Japan-Russia
Research Cooperative Program of Japan Society for the Promotion of
Science.  The studies are partially supported by the Russian Foundation
Base Research (RFBR) grant N~05-02-17556 and the mutual RFBR and Japan
Society for the Promotion of Science (JSPS) grant N~05-02-19710. The JVN
project is led by the National Astronomical Observatory of Japan~(NAOJ)
that is a branch of the National Institutes of Natural Sciences~(NINS),
Hokkaido University, Gifu University, Yamaguchi University, and
Kagoshima University, in cooperation with the Geographical Survey
Institute~(GSI), the Japan Aerospace Exploration Agency~(JAXA), and the
National Institute of Information and Communications Technology~(NICT).
TK is supported by a 21st Century COE Program at
Tokyo Tech ``Nanometer-Scale Quantum Physics'' by the
Ministry of Education, Culture, Sports, Science and Technology.

% Table 1
\begin{longtable}{lrrrrrr}
\caption{Flux densities of Cyg X-3}
\label{tbl:flux}       % Give a unique label
\hline
\hline
MJD & $S_{\mbox{8.4 GHz}}$&$S_{\mbox{23 GHz}}$&$S_{\mbox{43 GHz}}$
 &$S_{\mbox{86 GHz}}$&$S_{\mbox{98 GHz}}$& $S_{\mbox{110 GHz}}$\\
$[day]$ &$[mJy]$&$[mJy]$&$[mJy]$&$[mJy]$&$[mJy]$&$[mJy]$\\
\hline
\endhead
\hline
\endfoot
\hline
\endlastfoot
53762.181 &                &       &       &       &12    &12    \\
53763.125 &      &20     &       &       &10    &9.6   \\
53764.083 &      &21     &       &       &12    &12    \\
53764.208 &      &       &15     &       &      &      \\
53768.125 &      &57     &58     &       &417   &442   \\
53768.292 &100   &       &       &       &      &      \\
53769.167 &      &       &       &       &3247  &1280  \\
53769.208 &1100  &896    &549    &338    &468   &408   \\
53770.042 &      &339    &       &       &      &      \\
53770.208 &650   &222    &       &       &85    &80    \\
53771.042 &      &224    &       &       &      &      \\
53771.167 &200   &       &       &       &89    &90    \\
53772.042 &      &122    &       &       &      &      \\
53772.167 &      &       &       &       &288   &260   \\
53772.910 &      &20     &       &       &      &      \\
53772.924 &      &155    &       &       &      &      \\
53772.934 &      &183    &       &       &      &      \\
53772.951 &      &219    &       &       &      &      \\
53772.965 &      &163    &       &       &      &      \\
53772.982 &      &156    &       &       &      &      \\
53772.995 &      &214    &       &       &      &      \\
53773.208 &      &       &       &       &104   &60    \\
53774.042 &      &49     &       &       &39    &35    \\
53774.208 &      &255    &       &       &      &      \\
53774.903 &      &922    &       &       &      &      \\
53774.924 &      &1080   &       &       &      &      \\
53774.937 &      &1365   &       &       &      &      \\
53775.024 &      &1732   &       &       &      &      \\
53775.042 &      &1589   &       &       &      &      \\
53775.096 &      &       &       &       &890   &850   \\
53775.104 &      &       &       &       &650   &620   \\
53775.115 &      &       &       &       &450   &340   \\
53775.125 &      &       &       &       &300   &250   \\
53775.135 &      &       &       &       &300   &300   \\
53775.140 &690   &       &       &       &      &      \\
53775.146 &      &       &       &       &370   &300   \\
53775.156 &      &       &       &       &320   &300   \\
53775.167 &      &       &       &       &590   &550   \\
53775.200 &710   &       &       &       &      &      \\
53775.208 &      &801    &587    &352    &      &      \\
53775.260 &650   &       &       &       &      &      \\
53775.320 &500   &       &       &       &      &      \\
53776.152 &      &52     &       &       &18    &12    \\
53776.958 &      &25     &       &       &      &      \\
53777.188 &      &74     &       &       &      &      \\
53778.208 &      &33     &       &       &      &      \\
53778.948 &      &34     &       &       &      &      \\
53779.208 &      &29     &       &       &      &      \\
53779.937 &      &18     &       &       &      &      \\
53787.125 &60    &       &       &       &      &      \\
53800.208 &1200  &       &       &       &      &      \\
53805.118 &      &2424   &       &       &      &      \\
53805.729 &      &1762   &       &       &      &      \\
53805.750 &      &1609   &       &       &      &      \\
53805.757 &      &       &1402   &1203   &      &      \\
53806.090 &      &805    &       &       &      &      \\
53806.104 &      &896    &       &       &      &      \\
53806.118 &      &733    &       &       &      &      \\
53806.139 &      &601    &       &       &      &      \\
53806.708 &      &764    &       &       &      &      \\
53807.771 &      &1192   &       &       &      &      \\
53808.771 &      &1782   &       &       &      &      \\
53813.000 &1530  &       &       &       &      &      \\
53813.083 &1260  &       &       &       &      &      \\
53866.014 &10000 &       &       &       &      &      \\
53866.847 &      &       &       &       &6000  &5200  \\
53867.847 &      &       &       &       &2200  &1400  \\
53868.847 &      &       &       &       &3000  &2500  \\
53869.847 &      &       &       &       &3900  &3500  \\
53872.014 &1500  &       &       &       &      &      \\
53883.847 &      &       &       &       &400	&400   \\
\end{longtable}


\begin{thebibliography}{}
\bibitem[Braes \& Miley(1972)]{braes1972}
  Braes, L. L. E., \& Miley, G. K.
  \ 1972, Nature, 237, 506

\bibitem[Doi et al.(2006a)]{doi2006a} Doi, A., et~al.\ 2006a, astro-ph/0612528 

\bibitem[Doi et al.(2006b)]{doi2006b} Doi, A., et~al.\ 2006b, \pasj, 58, 777 

\bibitem[Fujisawa et al.(2007)]{fujisawa2007} Fujisawa, K., et~al.\ 2007, \ in prep.

\bibitem[Gregory et al.(1972)]{gregory1972}
  Gregory, P. C., Kronberg, P. P., Seaquist, E. R., Hughes, V., A.,
  Woodsworth, A., Viner, M. R., \& Retallack, D.
  \ 1972, Nature, 239, 440

\bibitem[Hjellming et al.(1974)]{hjellming1974} Hjellming, R. M., Brown,
			      R. L., \&~Blankenship, L. C. 1974, \apj,
			      194, L13

\bibitem[Hjellming \&~Johnston(1988)]{hjellming1988}
  Hjellming, R. M., \&~Johnston, K. J.
  \ 1988, ApJ, 328, 600

\bibitem[Kobayashi et al.(2003)]{kobayashi2003} Kobayashi, H., et~al.\ 2003, Astronomical Society of the Pacific Conference Series, 306, 367 

\bibitem[Greisen(2003)]{greisen2003} Greisen, E.~W.\ 2003, Information Handling in Astronomy - Historical Vistas, 109 

\bibitem[McCollough et~al.(1999)]{mccollough1999}
  McCollough, M.~L., Robinson, C.~R., Zhang, S.~N., Harmon, B.~A.,
  Hjellming, R.~M., Waltman, E.~B., Foster, R.~S., Ghigo, F.~D.,
  Briggs, M.~S., Pendleton, G.~N., \&~Johnston, K.~J.
  \ 1999, \apj, 517, 951

\bibitem[Miller-Jones et~al.(2004)]{miller-jones2004} Miller-Jones, J.~C.~A., Blundell, K.~M., Rupen, M.~P., Mioduszewski, A.~J., Duffy, P., \& Beasley, A.~J.\ 2004, \apj, 600, 368 

\bibitem[Okumura et~al.(2000)]{okumura2000} Okumura, S.~K., Momose, M., Kawaguchi, N. et~al., \ 2000, PASJ, 52, 393

\bibitem[Ott et al.(1994)]{ott1994} Ott, M., Witzel, A., Quirrenbach, A., et al., 1994, A\&A   284, 331

\bibitem[Predehl et~al.(2000)]{predehl2000}
  Predehl, P., Bruwitz, V., Paerels, F., \&~Tr\"umper, J.
  \ 2000, A\&A, 357, L25

\bibitem[Schalinski et~al.(1995)]{schalinski1995} Schalinski, C.~J., et~al.\ 1995, \apj, 447, 752

\bibitem[Schalinski et~al.(1998)]{schalinski1998}
  Schalinski, C.~J., Johnston, K.~J., Witzel, A., Waltman, E.~B.,
  Umana, G., Pavelin, P.~E., Ghigo, F.~D., Venturi, T.,
  Mantovani, F., Foley, A.~R., Spencer, R.~E., \&~Davis, R.~J.
  \ 1998, A\&A, 329, 504

\bibitem[Shepherd(1997)]{shepherd1997} Shepherd, M.~C.\ 1997, ASP Conf.~Ser.~125: Astronomical Data Analysis Software and Systems VI, 125, 77 

\bibitem[Shibata et~al.(1998)]{shibata1998} Shibata, K.~M., Kameno, S., Inoue, M., \&~Kobayashi, H.\ 1998, ASP Conf.~Ser.~144: IAU Colloq.~164: Radio Emission from Galactic and Extragalactic Compact Sources, 144, 413

\bibitem[Trushkin et~al.(2006)]{trushkin2006}
  Trushkin, S.~A., Nizhelskij, N.~A., Bursov, N.~N., \&~Majorova, E.~K.
  \ 2006, in IAU Symp.\ 238, 166

\bibitem[Tsuboi et~al.(2006)]{tsuboi2006} Tsuboi, M. et~al.\ 2006, 
ATel, 727, 1 

\bibitem[Tosaki et~al.(2006)]{tosaki2006} Tosaki, T., Tsuboi, M.
Nakanishi, K., Trushkin, S., Fujisawa, K., Kameya,O., Kotani, T., \&~Kawai, N.\ 2006, 
ATel, 952, 1 

\bibitem[Tsutsumi, Morita, \&~Umeyama(1997)]{tsutsumi1997} Tsutsumi, T., Morita, K.-I., \&~Umeyama, S. \
1997, in ASP Conf. Ser. 125, Astronomical Data Analysis Software and
Systems VI, ed. G. Hunt \& H. E. Payne (San Francisco: ASP), 50

\bibitem[Waltman et~al.(1994)]{waltman1994}
  Waltman, E.~B., Fiedler, R. L., Johnston, K. L., \&~Ghigo, F.~D.
  \ 1994, AJ, 108, 179


\end{thebibliography}
\end{document}